# Secure positioning and non-local correlations

*Muhammad Nadeem*
*Department of Basic Sciences,*
*School of Electrical Engineering and Computer Science*
*National University of Sciences and Technology (NUST)*
*H-12 Islamabad, Pakistan*
*muhammad.nadeem@seecs.edu.pk*

Recently, the problem of secure position-verification has been extensively analyzed in a formal notion where distant verifiers send encrypted challenge along with the decryption information to the prover. However, currently it is known that all the existing position-verification scheme are insecure and secure positioning is impossible [*Nature* **479**, 307-308 (2011)]. We propose here a different notion for position-verification where distant verifiers determine the actions of the prover through non-local correlations generated by local measurements at the provers' site. The proposed scheme guarantees secure positioning even if the verifiers do not have any pre-shared data with the prover; position is the only credential of the prover. Our proposed scheme is secure in general and evades known quantum attacks based on instantaneous non-local computations in particular. The scheme enables verifiers to identify dishonest provers with very high probability $\rho \geq 1 - 1/2^n$, where n is the number of entangled pairs used.

**Introduction**

In a general position-verification scheme, a prover located at a specified position convinces a set of verifiers at distant reference stations that he/she is indeed at the specific position. In a formal notion of position-verification, different verifiers send a secret message and a key to decrypt that message in pieces to the prover. That is, each verifier sends a bit of key such that all the key bits and the message arrive at the position of the prover concurrently. If the prover decrypts the message correctly and sends the result to all verifiers in time, position-verification scheme enables the verifiers to verify his position jointly. But if one or a set of dishonest provers, not at the specified position, intercept the communication and try to convince verifiers that they are at the specified position, a secure position-verification scheme enables the verifiers to reject it with high probability.

An unconditionally secure position-verification is impossible in classical cryptography where classical data can be copied [1]. A large number of quantum position-verification schemes [2-8] have also been proposed but unfortunately all these schemes are proved to be insecure later. Currently it is known that if the position of the prover is his/her only credential and he/she does not have any pre-shared data with the verifiers, then unconditional position-verification is impossible [8,9]. The security of any quantum position-verification (QPV) scheme can be destroyed by coalition of dishonest provers through teleporting quantum states back and forth and performing instantaneous non-local quantum computation, an idea introduced by Vaidman [10]. S. Beigi and R. Konig showed that if dishonest provers posses an exponential (in n) amount of entanglement then they can successfully attack any QPV scheme where n qubits are communicated [11]. Burrman *et al* have also shown that the minimum amount of entanglement needed to perform a successful attack on QPV schemes must be at least linear in the number of communicated qubits [9,12].

In the search of unconditional security, some authors proposed that secure positioning is possible if the prover and the verifiers pre-share some classical or quantum data. Kent proposed that secure quantum position-verification is possible if the prover and one of the verifiers pre-share some classical bit string unknown to eavesdroppers [13]. This secret data can then be used as a secret key to authenticate the communication. Recently, we have also shown that key-based quantum position verification can be securely achieved if distant verifiers and the prover pre-share entangled states [14]. The verifiers and the prover obtain secret keys through entanglement swapping [15] and then use these keys for authentication of secret messages.

Recently, we showed that combination of causality and quantum non-locality promises fascinating advancement in getting unconditional security from dishonest users [16,17]. For example quantum commitment scheme [16] and general framework for relativistic quantum cryptography [17] that implements important tasks such as oblivious transfer, two-sided two-party secure computation, and ideal coin tossing and unconditionally secure bit commitment.

We propose here a scheme for secure positioning based on causality and quantum non-locality. The distant verifiers neither send a secret key to the prover for decrypting the challenge nor do they have any pre-shared data with the prover. The verifiers determine the actions of the prover through non-local correlations generated by local Bell measurements from a specific position. Teleportation [18] is used to insure that any group of dishonest provers, not at the position to be verified, cannot simulate their actions with the prover who is supposed to be at the specified position. Proposed scheme guarantees secure positioning by bounding the prover to receive, measure (in publically known basis) and teleport quantum information simultaneously as do the verifiers.

**Teleportation**
Teleportation is the most important step in our proposed scheme for secure positioning. In general teleportation works as follows: Suppose Alice and Bob share a maximally entangled state in Bell basis

$$|ab\rangle = \frac{|0\rangle|b\rangle + (-1)^a |1\rangle|1 \oplus b\rangle}{\sqrt{2}} \tag{1}$$

where $a,b \in \{0,1\}$ and $\oplus$ denotes addition with mod 2. Bob can send an arbitrary quantum state $|\psi\rangle = \alpha|0\rangle + \beta|1\rangle$ to Alice instantly by performing Bell state measurement [19] (BSM) on $|\psi\rangle$ and his half $|b\rangle$ of entangled pair. If Bob gets classical 2-bit string $bb'$, Alice's entangled half $|a\rangle$ instantly becomes one of the four possibilities:

$$|\psi'\rangle = \sigma_z^k \sigma_x^{k'} |\psi\rangle \tag{2}$$

where $k$ and $k'$ depend on the Bell state $|ab\rangle$ shared between them. For example, if they share a Bell state $|00\rangle$ then $k=b$ and $k'=b'$. If shared Bell state is $|01\rangle$ then $k=b$ and $k'=1 \oplus b'$. If they share Bell state $|10\rangle$ then $k=1 \oplus b$ and $k'=b'$ while for $|11\rangle$, $k=1 \oplus b$ and $k'=1 \oplus b'$. If Bob sends two classical bits $bb'$ to Alice, she can easily recover $|\psi\rangle$ by applying suitable unitary operators. However, without knowing shared entangled state $|ab\rangle$ or BSM result $bb'$ of Bob, $|\psi'\rangle$ remains totally random to Alice and we use this fact in our scheme for secure positioning.

Secure positioning & non-local correlations

**Quantum scheme for secure positioning**

We assume that the sites of the prover and reference stations are secure from adversary; enabling them to store and hide the quantum data and process. We also assume that the verifiers can communicate both classical and quantum information securely with each other. However, all the quantum/classical channels between verifier(s) and the prover are insecure. Moreover, there is no bound on storage, computing, receiving and transmitting powers of dishonest provers. They can interfere or jam communication of the prover without being detected. In short, dishonest provers have full control of environment except position of the prover and reference stations.

All reference stations and the prover have fixed positions in Minkowski space-time. Both quantum and classical signals can be sent between prover and verifiers at the speed of light while the time for information processing at position of the prover and reference stations is negligible. For simplicity, we discuss our scheme for two verifiers $V_1$ and $V_2$ at distant reference stations collinear with prover P, such that the prover is at a distance $x$ from both reference stations. Explicit procedure of our quantum scheme for secure positioning is described below where both verifiers and the prover measure and send quantum states in publically known Hadamard basis.

1). At time t=0, verifiers $V_1$ and $V_2$ secretly prepare EPR pairs $|v_1 p_1\rangle \in H_{V_1} \otimes H_{P_1}$ and $|v_2 p_2\rangle \in H_{V_2} \otimes H_{P_2}$ respectively and each sends second half to P.

2). At time t=$x$/c, $V_1$ teleports state $|\psi\rangle = |\pm\rangle$ to P. As a result $V_1$ gets classical information $vv' \in \{00,01,10,11\}$ while the P's half $|p_1\rangle$ becomes $|\psi'\rangle = \sigma_z^k \sigma_x^{k'} |\psi\rangle$ where values of $k$ and $k'$ depend on $vv'$ and $|v_1 p_1\rangle$ only known to $V_1$.

3). At the same time t=$x$/c, P measure $|p_1\rangle$ in agreed Hadamard basis and teleports $|\psi'\rangle$ to $V_2$ over EPR channel $|v_2 p_2\rangle$. Entangled half $|v_2\rangle$ in possession of $V_2$ becomes $|\psi''\rangle = \sigma_z^l \sigma_x^{l'} |\psi'\rangle$ where values of $l$ and $l'$ depend on P's BSM result $pp' \in \{00,01,10,11\}$ and $|v_2 p_2\rangle$. Simultaneously, P sends classical 2-bit string $pp'$ and quantum state $|\psi'\rangle$ to both $V_1$ and $V_2$.

4). At time t=2$x$/c, verifier $V_2$ verifies that whether $|\psi'\rangle$ and $|\psi''\rangle$ are consistent with BSM result $pp'$ of P or not. Similarly $V_1$ validates whether $|\psi\rangle$ and $|\psi'\rangle$ are consistent with his BSM result $vv'$ or not. If both $V_1$ and $V_2$ receive verified information from P, they exchange their measurement outcomes somewhere in their causal future and verify the position of P if P has replied within allocated time; at t=2$x$/c.

In the proposed scheme, classical communication between the prover and verifiers can be reduced to single bit only. Since $|\psi\rangle = |\pm\rangle$, Pauli encodings $\sigma_z^l \sigma_x^{l'} \in \{I, \sigma_x\}$ give same outcome $|\psi''\rangle$ at $V_2$ site up to overall phase factor. Similarly, $\sigma_z^l \sigma_x^{l'} \in \{\sigma_z, \sigma_z \sigma_x\}$ give same outcome $|\psi''\rangle$. For example, if $|v_2 p_2\rangle \in \{|00\rangle, |01\rangle\}$, then P's BSM result $pp' \in \{00,01\}$ will result in $\sigma_z^l \sigma_x^{l'} \in \{I, \sigma_x\}$ while that of $pp' \in \{10,11\}$ will generate $\sigma_z^l \sigma_x^{l'} \in \{\sigma_z, \sigma_z \sigma_x\}$. Hence, instead of sending classical 2-bit string $pp'$ to verifiers, prover can simply announce that whether first bit of his BSM result $pp'$ is $p=0$ or $p=1$. In short, if the verifiers run this scheme with $H_{V_1} = (C^2)^{\otimes n}$, $H_{V_2} = (C^2)^{\otimes n}$ and $H_P = H_{P_1} \otimes H_{P_2} = (C^2)^{\otimes n} \otimes (C^2)^{\otimes n}$, the proposed scheme enables them to identify dishonest provers with very high probability; $\rho \geq 1 - 1/2^n$.

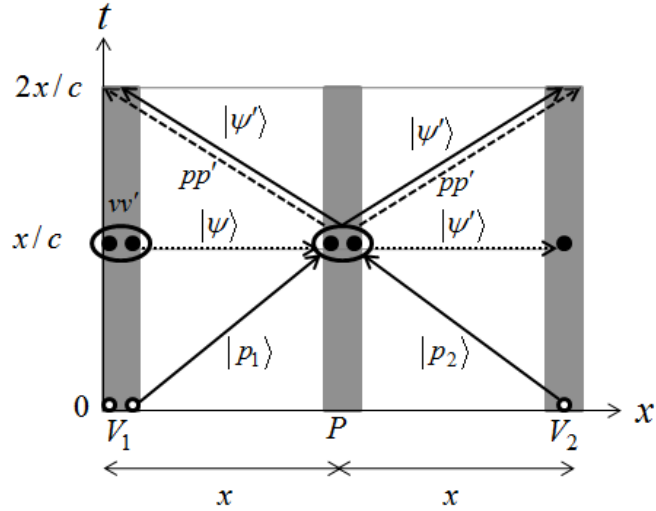

**Figure 1:** Quantum scheme for secure positioning: Solid arrows represent quantum states; dotted arrows show teleportation from $V_1$ to P and P to $V_2$ respectively while dashed arrows show classical information from P to the verifiers.

**Security analysis**
In [8,9,11,12], it was shown that no position-verification scheme can evade quantum attacks based on instantaneous non-local quantum computations by dishonest provers. For example, the impossibility proof discussed by Buhrman *et al* are applicable to the QPV schemes where the prover P receives a quantum system $H_{P_1}$ from the verifier $V_1$ and a system $H_{P_2}$ from $V_2$; $H_{P_1}$ and $H_{P_2}$ are components of some larger quantum system $H = H_{P_1} \otimes H_{P_2} \otimes H_{V_1} \otimes H_{V_2}$. The prover then applies some unitary transformations $U$ on $H_P = H_{P_1} \otimes H_{P_2}$ depending on the classical information $\mathcal{V}_1$ and $\mathcal{V}_2$ obtained from $V_1$ and $V_2$ respectively. In this existing notion, the verifiers validate the exact position of the prover P if he replies correct information $U(H_P)$, consistent with $\mathcal{V}_1$ and $\mathcal{V}_2$ and hence larger quantum system $H$, within allocated time.

Such position-verification schemes are not secure since a group of dishonest provers $P_1$ and $P_2$ at different positions from that of P can obtain both quantum ($H_{P_1}$ and $H_{P_2}$) and classical information ($\mathcal{V}_1$ and $\mathcal{V}_2$) before the prover P. Hence they can reply exact information $U(H_P)$ to the verifiers within time by performing non-local instantaneous computations. As a result, verifiers cannot differentiate between the responses from the prover P and group of dishonest provers at positions different from that of P.

Here we show that our proposed scheme is secure against such entanglement-based attacks: verifiers neither send qubit-wise encrypted quantum systems as a challenge nor classical information for decrypting that challenge. Instead, verifiers determine the actions of the prover through non-local correlations generated by local measurements from a specific position. The verifiers starts the scheme at t=0 by preparing quantum system $H = H_{V_1 P_1} \otimes H_{V_2 P_2}$ where $H_{V_1 P_1} = H_{V_1} \otimes H_{P_1}$ is a maximally entangled system shared between $V_1$ and P while $V_2$ and P share entangled system $H_{V_2 P_2} = H_{V_2} \otimes H_{P_2}$ at time t=x/c. Verifiers control the space time position where they want to reveal the challenge through teleportation. Before that space time position (occupied by prover P), dishonest provers cannot extract required information from quantum

Secure positioning & non-local correlations systems $H_{P_1}$ and $H_{P_2}$ in the causal past of prover P since these quantum systems do not contain information required for position-verification.

Suppose dishonest prover $P_1$ is between $V_1$ and P at position $(x-\delta,0)$ while $P_2$ is between $V_2$ and P at position $(x+\delta,0)$ respectively. Now $P_1$ can intercept $H_{P_1}$ and get entangled with the verifier $V_1$ in a state $H_{V_1P_1}$ while $P_2$ shares entangled state $H_{V_2P_2}$ with verifier $V_2$ at time $t=(x-\delta)/c$.

In our proposed scheme, prover P (or dishonest provers) has to reply with both quantum state $|\psi'\rangle$ and classical 2-bit string $pp'$ simultaneously. In other words, P (or dishonest provers) has to receive teleported state $|\psi'\rangle$ from $V_1$ and then teleport same state $|\psi'\rangle$ to $V_2$. Since verifier $V_1$ knows the definite state $|\psi'\rangle = \sigma_z^k \sigma_x^{k'} |\psi\rangle$ from initially prepared EPR pair $|v_1 p_1\rangle$ and his BSM result $vv'$, hence verifiers $V_1$ and $V_2$ can verify that whether the announced BSM result from prover P (or dishonest provers) is consistent with $|\psi''\rangle = \sigma_z^l \sigma_x^{l'} |\psi'\rangle$ or not.

Moreover, verifier $V_1$ teleports quantum state $|\psi\rangle$ over EPR pair $|v_1 p_1\rangle$ not before time $t=x/c$, hence specially separated dishonest provers $P_1$ and $P_2$ are restricted to perform only following non-local computation during time interval $\{(x-\delta)/c, x/c\}$: $P_1$ ($P_2$) performs entanglement swapping on $|v_1 p_1\rangle$ ($|v_2 p_2\rangle$) and pre-shared entangled state $|p_1 p_2\rangle$ with $P_2$ ($P_1$) such that only $P_2$ ($P_1$) is entangled with both verifiers $V_1$ and $V_2$. In that case $P_1$ (or $P_2$) can send both quantum state and classical BSM result to both $V_1$ and $V_2$ but not before $t=(2x+\delta)/c$.

Even if $P_1$ and $P_2$ have infinite amount of pre-shared entanglement and perform non-local instantaneous computations through multiple rounds of teleportation [10], $P_1$ and $P_2$ can agree on $|\psi'\rangle = \sigma_z^k \sigma_x^{k'} |\psi\rangle$ and required classical 2-bit string only at time $t=(x+2\delta)/c$ and hence can send required information to both $V_1$ and $V_2$ not before time $t=(2x+\delta)/c$.

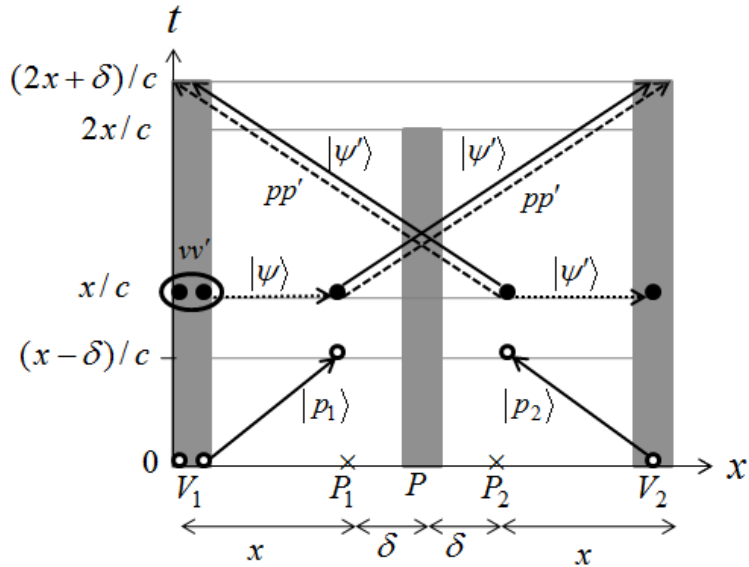

**Figure 2:** Quantum scheme for secure positioning with dishonest provers $P_1$ and $P_2$.

**Discussion**

We proposed here a different notion for secure positioning where distant verifiers do not send a secret key to the prover along with challenge but the actions of the honest prover are determined through non-local correlations obtained by local measurements at the provers' site. The causality principle insures that the proposed quantum position-verification scheme is secure against entanglement-based attacks even if eavesdroppers have infinite amount of pre-shared entanglement and power of non-local quantum measurements in negligible time.

The proposed scheme for secure-positioning can be efficiently and reliably implemented using existing quantum technologies. Since the quantum memory for reliable storage of entangled quantum systems is not available yet, we use more practical setup where the prover and verifiers can store and communicate classical information only.

The combination of causality and quantum non-locality as discussed here for secure positioning promises fascinating advancement in getting unconditional security from dishonest users. For example, the receiver can trust the information he receives only if the scheme verifies position of the sender and validates sender's actions in a single round. This bounds sender to reveal valid information within allocated time and guarantees to the sender that the receiver will not be able to get information unless sender reveals.

Hopefully, proposed scheme for secure positioning would help in achieving quantum cryptographic tasks that are considered to be impossible otherwise. In the broader perspective, this notion for secure positioning could be useful to understand relativistic quantum theory on the basis of quantum information science.